# Optimization of Neon Soft X-ray Emission in Low Energy Dense Plasma Focus Device

M. A. Malek[1,3], M. K. Islam[2], M. Salahuddin[1]

[1]Jahangirnagar University, Savar, Dhaka, Bangladesh
[2]Plasma Physics Division, Atomic Energy Center, Dhaka, Bangladesh
[3]Green University, Dhaka, Bangladesh
E-mail: malekphy@gmail.com

**Abstract:** The Lee model code is used in numerical experiments for characterizing and optimizing neon soft X-ray ($Y_{sxr}$) yield of UNU/ICTP PFF machine operated at 14 kV and 30 µF. The neon $Y_{sxr}$ yield of the dense plasma focus device is enhanced by reducing static inductance ($L_0$) and anode length ($z_0$) along with increasing anode radius ($a$) and cathode radius ($b$), keeping their ratio ($c = b/a$) constant at 3.368. At the optimum combination of the electrodes geometry and static inductance, the maximum computed value of neon $Y_{sxr}$ yield is 63.61 J at operating pressure 3.3 Torr with corresponding X-ray yield efficiency 2.16%, while the end axial speed becomes 6.42 cm/µs. This value of neon $Y_{sxr}$ yield is twelve to thirteen times higher than the measured value (5.4 ± 1 J) at 3.0 Torr. It is also found that this neon $Y_{sxr}$ yield is improved around seven times from previously computed value (9.5 J) at 3.5 Torr for optimum anode configuration of this machine. Our obtained results of neon $Y_{sxr}$ yield are also compared with the computed results of AECS-PF2 machine operated at 15 kV and 25 µF and is found that our results are about three times better than that from the optimized AECS-PF2 at $L_0$ = 15 nH.

## 1. Introduction

The dense plasma focus (DPF) device is a non-radioactive co-axial accelerator with relatively simple operating principle that produces a high-density, high-temperature plasma along with pulsed fusion neutron yield, soft and hard X-rays, high energy electron and ion beam, and electromagnetic waves [1,2,3]. This device is easy to construct, requires minimum maintenance and cost. The pulsed X-ray emitted from it is the highest among all other existing devices of equivalent operating energy [4]. The DPF device as high intensity pulsed X-ray source has a wide range of real life applications such as: X-ray spectroscopy [5], X-ray microscopy and lithography [6], X-ray laser pumping [7], X-ray crystallography [8], X-ray radiography [9], X-ray back-lighter [10] and X-ray micromachining [11]. The United Nations University/International Center for Theoretical Physics Plasma Focus Facility (UNU/ICTP PFF) is a 3.3 kJ Mather-type DPF machine which is switched on by a parallel-plate switching cascade air-gap, powered by 15 kV and 30 μF Maxwell Capacitor [12]. The UNU/ICTP PFF machine has a unique contribution of plasma focus research. The UNU, ICTP and AAAPT (Asian African Association for Plasma Training) developed this device to initiate and promote practical knowledge and skills in plasma physics as well as fusion, in developing countries [13]. This machine produces realistic focusing action operating in several gases (He, Ne, Ar, $H_2$, $CO_2$, $D_2$, $N_2$ etc.) [14]. The neon $Y_{sxr}$ yield for optimized DPF machine with operating energy from 0.2 kJ to 1 MJ was computed through numerical experiments by Lee model code and observed that the Neon is to be a suitable operating gas for the device as the source of soft X-ray yield [15].

The Lee model code is used to compute the realistic focus parameters along with the soft X-ray yield by only adjusting the computed discharge current wave trace with the experimentally measured one. In the case of NX2 DPF machine, this code has been successfully used showing a reasonable good agreement between the computed and measured values of neon $Y_{sxr}$ yield as a function of pressure [4]. Therefore, the Lee model code is used to compute and optimize a DPF machine for improving the realistic $Y_{sxr}$ yield.

For enhancing X-ray yields from the device, many efforts have been made by changing the bank, tube and operational parameters such as: energy of the bank static, circuit inductance, discharge current, electrode configuration (shape and materials) insulator materials and dimensions, gas composition, and filling gas pressure [16].

The measured value of neon $Y_{sxr}$ yield from UNU/ICTP PFF was (5.4 ± 1) J at optimum pressure of 3.0 Torr with corresponding efficiency of 0.18% [2]. The numerical experiment on this device was carried out using Lee model code to compute the optimum neon $Y_{sxr}$ yield keeping cathode radius '$b$' fixed at 3.2 cm, the anode length '$z_0$' was drastically decreased from 16 to 7 cm, whilst the anode radius '$a$' was slightly increased from 0.95 to 1.2 cm from the standard configuration. As a result, the neon $Y_{sxr}$ yield increased to 9.5 J at optimum pressure 3.5 Torr with corresponding efficiency of 0.32% [13]. This computed efficiency of $Y_{sxr}$ was improved two to three times from the experimental value (0.18%) of the present UNU/ICTP PFF.

The AECS-PF2 is a 2.8 kJ Mather-type DPF machine operated with 15 kV and 25 µF capacitor. Using the Lee model code, the reduction effect of static inductance ($L_0$) on neon $Y_{sxr}$ yield from this device was investigated and optimized with electrodes geometry for maximum neon $Y_{sxr}$ yield. At optimum configuration of AECS-PF2, the $L_0$ was reduced from 280 to 15 nH and hence the obtained neon $Y_{sxr}$ yield was improved from its typical value 0.04 to 21 J at operating pressure of 2.8 Torr with corresponding efficiency of about 0.77% [16].

It is understood that the effect of $L_0$ on neon $Y_{sxr}$ yield in optimization of UNU/ICTP PFF has not been done. Therefore, this device is optimized using Lee model code to enhance neon soft X-ray yield finding the optimum $L_0$ with its corresponding combination of electrode geometry ($z_0$, '$a$' & '$b$').

In this paper, section 2 describes the Lee model code, section 3 describes the method of numerical experiment and the detailed descriptions of numerical experiment on UNU/ICTP PFF with neon gas is given in section 4. The section 5 explains the finding process of optimum static inductance with its corresponding electrode configuration for possible maximum neon soft X-ray yield in optimized UNU/ICTP PFF.

## 2. Lee Model Code

The electrical circuit and plasma focus dynamics, thermodynamics, radiations are coupled by 'Lee model code' which enables a realistic simulation that analyse all of the gross properties and performances of a DPF machine [17]. This code is used in the interpretation of experiments and design of a DPF [18]. An improved 5-phase code incorporating finite small disturbance speed, radiation and radiation-coupled dynamics was used [4] and was first web-published [19] in 2000. Plasma self-absorption was included [17, 19] in 2007 improving soft X-ray simulation in neon, argon and xenon among other gases. It has been widely used as a

Page **3** of **21**

complementary facility in several machines, such as: UNU/ICTP PFF [4, 12], NX1 and NX2 [4], DENA [19]. It has also been used in other machines for design and interpretation including sub-kJ DPF machines [20], FNII [21], the UBA hard X-ray source [22], KSU PF [23] and a cascading DPF [24]. Computed information from Lee model code incudes: axial and radial speeds and dynamics [12, 23], focus pinch duration and dimensions, average pinch temperatures and densities, soft X-ray characteristics and yield [4, 13], optimization of machines [4, 12, 13, 17], and adaptation with Modified Lee (ML) model for Filipov-type DPF devices [19]. The reason of continuous development of the code for the last three decades is that there is no breakthrough to publish the basis and description of the model code; although many details in recent years, as they evolved, the Plasma Focus Studies Institute's website has been described [17]. The modified 6-phase version of the Lee model code RADPFV6.1b for Type-2 (high inductance DPF) machines which have been found to be incompletely fitted with the 5-phase model due to a dominant anomalous resistance phase [25]. The soft X-ray emission is calculated by subtracting the plasma self-absorption from the generated soft X-ray energy (line radiation), which mainly depends on the temperature and density. In the Lee model code, this effect is included for obtaining neon $Y_{sxr}$ yield [17]. This effect is caused by the reabsorption of an emitted photon (in this case, the soft X-ray) from an atom/ion by another plasma component before escaping the plasma region, resulting in a lower radiation yield.

In the Lee model code, the rate of neon line radiation is calculated as follows [17]:

$$\frac{dQ_L}{dt} = -4.6 \times 10^{-31} n_i^2 Z Z_n^4 (\pi r_p^2) z_f / T \quad \text{………………………………………..………...…} (1)$$

where, $Q_L$ is the neon line radiation, $Z_n$ is the atomic number, $Z$ is the effective charge number, $n_i$ is the number density, $r_p$ the pinch radius, $z_f$ is the pinch column length and $T$ is the average temperature of pinch plasma. In the calculation of the code, $Q_L$ is computed by integrating over the pinch duration. The neon soft X-ray yield is to be equivalent to the line radiation yield i.e., $Y_{sxr} = Q_L$ within the temperature range 200 – 500 eV ($2.32 \times 10^6$ to $5.8 \times 10^6$ K) which corresponds to an end axial speed 6 - 7 cm/µs [2] at the modified 6-phase Lee model code.

## 3. Method of Numerical Experiment

The measured discharge current waveform is a significant indicator to realistically simulate and analyze all the gross performance of any DPF device. Important information such as: the axial and radial phase dynamics, temperature and thermodynamic properties, the crucial energy transfer into the focus pinch that causes nuclear fusion and hence the radiation yields from the device is carried out from the current waveform [18]. This is why the discharge current trace fitting is one of the best important techniques to optimize and configure a DPF. Therefore, the fitting of the computed discharge current waveform to the measured value through numerical experiment using Lee model code provides a lot of valuable insights of the pinched plasma. First of all, the measured data of the corresponding total discharge current waveform is collected either from laboratory experiment or picked out from published article. To start numerical experiments, the Lee model code is configured for any DPF by providing the tube parameters: $z_0$, '$a$' and '$b$'; the bank parameters: $L_0$, $C_0$ and stray circuit resistance $r_0$ and the operational parameters: $V_0$, $P_0$ and the fill gas [1, 18]. Then the computed total discharge current waveform is fitted to the measured waveform by sequential adjustment of the four model parameters: mass swept-up factor ($f_m$), plasma current factor ($f_c$) in the axial phase and accordingly radial mass factor ($f_{mr}$), radial current factor ($f_{cr}$) in the radial phase. First initiative of fitting, the values of axial model parameters $f_m$ and $f_c$ are adjusted in such a manner that the rising slope of computed current trace and peak discharge current are reasonably agreed with the measured total current trace [1]. Then the radial model parameters $f_{mr}$, $f_{cr}$ are varied until the computed slope and the deep fit with the measured values.

## 4. Numerical Experiments on UNU/ICTP PFF with Neon Filling Gas

To start the numerical experiments, the Lee model code (version: RADPFV6.1b) is configured for the UNU/ICTP PFF device with the following published parameters:
Bank parameters: static inductance $L_0$ = 110 nH, $C_0$ = 30 µF and stray circuit resistance $r_0$ = 12 mΩ; Tube parameters: cathode radius $b$ = 3.2 cm, anode radius $a$ = 0.95 cm, and anode length $z_0$ = 16 cm; Operation parameters: voltage $V_0$ = 14 kV and pressure $P_0$ = 2.8 Torr neon gas [13].
A measured discharge current waveform of the UNU/ICTP PFF at 14 kV and 2.4 Torr neon filling gas has been collected from the reference [26]. To obtain a reasonably good fit of the numerically computed total discharge current waveform to the measured waveform (Fig. 1), the following model parameters has been obtained: $f_m$ = 0.05, $f_c$ = 0.7, $f_{mr}$ = 0.2 and $f_{cr}$ = 0.8 [13]. These fitted values of the model parameters have been used for the computation of all the discharges at pressures from 1.0 to 4.2 Torr, keeping fixed all of the mentioned above bank, tube and operation parameters [13]. From the computed results, it is noticed that the neon $Y_{sxr}$ yield increases with increasing gas pressure until it goes to the maximum value of about 3.92 J at $P_0$ = 3.3 Torr with corresponding efficiency of 0.13% after which it decreases with further rising of the pressure. At this optimum pressure, the end axial speed is $v_a$ = 5.4 cm/µs, the total peak discharge current is $I_{peak}$ = 180 kA, the pinch current is $I_{pinch}$ = 103 kA and the focusing time is about 3.97 µs. It is observed that the focusing time increases with increase in the gas pressure. This is because higher the gas pressure, lower the current sheath velocity in both axial and radial phases and hence focus time comes slower with increase in gas pressure.

The characteristics of the variation of neon $Y_{sxr}$ yield with pressure depend on two major factors. Firstly: at the optimum pressure, the present configuration of UNU/ICTP PFF generate the appropriate end axial velocity of about 5.4 cm/µs, which correspond to the pinch temperature of $2.04 \times 10^6$ K, which is very close to the correct pinch temperature range for neon gas [27]. Secondly: the radiation yield is proportional to the square of the plasma density. Therefore, when the pressure is increased from a low value, the density of the pinched radiating plasma increases as a result the X-ray emission increases. Thus, at very low pressure the pinch plasma density is too low whilst the pinch temperature is to be very high due to high current sheath velocity. On the other hand, at very high pressure the pinch plasma density would be high and the corresponding pinch temperature may be too low for low current sheath velocity. In both cases, the pinch temperature may be away from the temperature range and hence the emitted neon $Y_{sxr}$ yield is low. Therefore, there would be an optimum pressure at which the pinch temperature and the corresponding end axial velocity are within the expected range whilst the density is still high enough for getting maximum neon $Y_{sxr}$ yield as shown in Fig. 2.

The measured values of neon $Y_{sxr}$ yield from UNU/ICTP PFF have been obtained by Liu [2] using a five-channel p-i-n soft X-ray detector confirmed by a calorimeter. In this experiment, the maximum value of neon $Y_{sxr}$ yield from this device was found to be about (5.4 ± 1) J at the optimum pressure of $P_0$ = 3.0 Torr with corresponding efficiency of 0.18%. At this optimum pressure, the typical values of the end axial speed is $v_a$ = 5.7 cm/µs, the total peak discharge current is $I_{peak}$ = 180 kA, the pinch current is $I_{pinch}$ = 111 kA.

In addition, many numerical experiments have been carried out to observe the effect of applied voltage on neon $Y_{sxr}$ yield from UNU/ICTP PFF with pressure. The variation of neon $Y_{sxr}$ yields with pressure from this device are plotted at applied voltages of 12, 13, 14 and 15 kV as shown in Fig. 3. From this figure, it is seen that at optimum pressure, the neon $Y_{sxr}$ yield rises from 2.74 to 4.49 J with increasing the applied voltage from 12 to 15 kV. It is also noticed for all applied voltages, the general natures of the variation of neon $Y_{sxr}$ yield with pressure are almost the same.

The electrode geometry of UNU/ICTP PFF has also been optimized with pressure through numerical experiments using Lee model code, keeping $c = b/a$ constant at 3.4 for enhancing neon $Y_{sxr}$ yield. At this practical optimization, the anode length $z_0$ is shorten drastically from 16 to 7 cm, the anode radius '$a$' is increased from 0.95 to 1.2 cm and the cathode radius '$b$' is kept unchanged at its original value of 3.2 cm [13]. At this optimum configuration, the computed values of $v_a = 4.9$ cm/μs, $I_{peak} = 184$ kA, $I_{pinch} = 141$ kA and the neon $Y_{sxr}$ yield is about 9.5 J at optimum $P_0 = 3.5$ Torr with corresponding efficiency of 0.32%.

It is understood from the above observations that the neon $Y_{sxr}$ yield rises two to three times from the standard UNU/ICTP PFF anode to the optimum anode configuration and it may be caused due to the increase of pinch current from 103 to 141 kA. Therefore, it would be a suitable technique to enhance $Y_{sxr}$ yield from further increase in $I_{pinch}$.

## 5. Soft X-ray Yield Versus Inductance and Electrode Geometry

To compute the optimum $L_0$ with its corresponding combination of electrode geometry ($z_0$, '$a$' and '$b$') for maximum neon $Y_{sxr}$ yield from the optimized UNU/ICTP PFF, the values of $C_0 = 30$ μF, $V_0 = 14$ kV and $P_0 = 3.3$ Torr neon and also the model parameters are kept constant throughout the numerical experiments. In these experiments, for each value of $L_0$ the corresponding $r_0$ is computed so that the factor RESF (RESF = stray circuit resistance/surge impedance = $r_0/\sqrt{(L_0/C_0)}$ remains fixed at 0.2. Also, for each $L_0$ the anode and cathode radii are adjusted in such a manner that their ratio $c = b/a$ is constant at 3.368. In our numerical experiment, the value of $L_0$ was varied from 100 nH to 1 nH using Lee model code.

In our numerical experiments he following techniques are applied to get the optimum combination of ($L_0$, $z_0$, '$a$' and '$b$') for maximum neon $Y_{sxr}$ yield [16]:

(i) The $P_0$ is kept constant at 3.3 Torr for all values of $L_0$ and also the value of $z_0$ is fixed at a certain value with each value of $L_0$.

(ii) Then the '$a$' and correspondingly '$b$' are varied keeping $c = 3.368$, until the maximum neon $Y_{sxr}$ yield is computed for the certain value of $z_0$.

(iii) The '$a$' as well as '$b$' are varied with different values of $z_0$ at each $L_0$ to obtain the optimum combination for maximum neon $Y_{sxr}$ yield.

(iv) After that another value of $z_0$ is chosen, the maximum neon $Y_{sxr}$ yield is computed by varying 'a' and 'b' and so on, until we have had the optimum combination of $z_0$, 'a' and 'b' for the best maximum neon $Y_{sxr}$ yield at a fixed value of $L_0$.

(v) The above procedures are repeated for gradually smaller $L_0$ until it was reached to 1 nH.

The reduction effects of $L_0$ on the discharge current waveforms with time are observed as shown in the Fig. 4. It is noticed from this figure that peak discharge current ($I_{peak}$) comes earlier for each reduction of $L_0$. For example: when $L_0$ = 30 nH, $I_{peak}$ = 318.45 kA at 1.28 µs, when $L_0$ = 20 nH, $I_{peak}$ = 368.11 kA at 1.02 µs, when $L_0$ = 10 nH, $I_{peak}$ = 453.91 kA at 0.60 µs, when $L_0$ = 5 nH, $I_{peak}$ = 508.44 kA at 0.31 µs and when $L_0$ = 3 nH, $I_{peak}$ = 623.21 kA at 0.28 µs. Therefore, the $z_0$ is needed to be reduced so that the time taken by the plasma current sheath to reach at the end of $z_0$ coincides with rising time of the $I_{peak}$ for maximum energy transfer to the crucial pinch region. At the same time, because of reducing $L_0$, the value of $I_{peak}$ increased as a result 'a' as well as 'b' were necessarily increased leading to longer pinch length ($z_{max}$) and hence a bigger pinch inductance ($L_p = \frac{\mu}{2\pi} \times ln\frac{b}{r_p} \times z_{max}$) is found [16].

Thus, the geometry of the machine moved from longer thinner (Mather-type) to shorter fatter (Filipov-type) one as shown in Fig. 5. From the careful observation of this figure it is seen that neon $Y_{sxr}$ yields start to increase proportionally from around $z_0$ = 4 cm to maximum at $z_0$ = 2.4 cm and then start to decrease sharply.

The values of 'a' and corresponding 'b' is varied with different values of $z_0$ at each $L_0$ to compute the optimum combination of them for getting maximum neon $Y_{sxr}$ yield, which corresponds closely to the largest $I_{pinch}$. The optimization of $z_0$, 'a' and 'b' with each $L_0$ for possible maximum value of neon $Y_{sxr}$ yield with corresponding efficiency (% of stored energy $E_0$ transfers into soft X-ray yield) is shown in Table 1. It is found from the table that $I_{peak}$ increases with each reduction in $L_0$ with no sign of any limitation as a function of $L_0$. Whereas, $I_{pinch}$ also rises gradually with reduction of $L_0$. Finally, $I_{pinch}$ reaches to a maximum value of 223.94 kA at $L_0$ = 10 nH whilst the neon $Y_{sxr}$ yield reaches to its maximum value beyond which $I_{pinch}$ as well as $Y_{sxr}$ start to decrease with further reduction in $L_0$, but the ratio of $I_{pinch}/I_{peak}$ drops progressively as $L_0$ decreases. Fig. 6 shows the limitation effect of $L_0$ on $I_{pinch}$ and $Y_{sxr}$ in UNU/ICTP PF operated with neon gas at 3.3 Torr, where $L_0$ is reduced from 100 nH to 1 nH. The following three reasons make the combined effect that limit the $I_{pinch}$ accordingly neon $Y_{sxr}$ yield at optimum combination of ($L_0$, $z_0$ 'a' and 'b'):

(i) If $L_0$ is reduced to zero then $I_{peak}$ would not be infinity because at $L_0 = 0$ though the serge impedance ($Z_0 = \sqrt{L_0/C_0}$) is zero, the dynamics of plasma current sheath produces an impedance, which then becomes the dominating load to limit the value of $I_{peak}$.

(ii) The capacitor bank will discharge within a short time through the focus pinch as $L_0$ is reduced to a very small value and it becomes more and more immediately coupled to the pinch.

(iii) The energy distributions and the requirement to adjust $z_0$ '$a$' as well as '$b$', the situation requires that as $L_0$ is decreased, the ratio of $I_{pinch}/I_{peak}$ reduces [28].

Looking at the table, it is also observed that as $L_0$ is reduced gradually and hence the corresponding '$a$' as well as '$b$' have to be increased, whereas $z_0$ decreased progressively to compute the optimum combination of ($L_0$, $z_0$ '$a$' and '$b$') for maximum neon $Y_{sxr}$ yield. In addition, for each reduction of $L_0$ with its corresponding optimum combination of electrodes geometry the plasma pinch dimensions (pinch radius $a_{min}$ and pinch length $z_{max}$) rise as a result the $Y_{sxr}$ yield increases. This variation of optimum neon $Y_{sxr}$ yield and its corresponding optimum combination of $z_0$, '$a$' and '$b$' with $L_0$ is shown in Fig. 7.

In the table, the maximum value of neon $Y_{sxr}$ yield with corresponding efficiency at optimum combination of static inductance and electrode geometry ($L_0$, $z_0$, '$a$' & '$b$') is represented by bold values. From this table, it is found that the maximum neon $Y_{sxr}$ yield is 63.61 J with corresponding efficiency of 2.16% at optimum electrode geometry of $z_0$ = 2.40 cm, '$a$ = 2 cm' and '$b$ = 6.736 cm' with $L_0$ = 10 nH in the optimized UNU/ICTP PFF at 14 kV and 3.3 Torr neon.

The neon $Y_{sxr}$ yield optimization for each value of $L_0$, varying $z_0$, '$a$' and '$b$' has also been observed in AECS-PF2 at 15 kV and 2.8 Torr neon [16]. It was computed the optimum static inductance and combination of electrode geometry as ($L_0$ = 15 nH, $z_0$ = 2.24 cm, '$a$ = 1.732 cm' and '$b$ = 5.83 = cm'), the maximum neon $Y_{sxr}$ yield was found to be about 21.77 J is shown in Fig. 8.

It is also found from Fig. 9 that the efficiency of soft X-ray emission from UNU/ICTP PFF increases with reducing $L_0$ and $z_0$. The efficiency is also higher for larger values of '$a$' and '$b$'. Accordingly, the maximum efficiency of neon $Y_{sxr}$ yield from optimized UNU/ICTP PFF is around 2.16% at the corresponding end axial speed of 6.42 cm/µs.

Whereas, the efficiency of neon $Y_{sxr}$ yield from AECS-PF2 at 15 kV and 2.8 Torr was found to be 0.77 % with the corresponding end axial speed of 5.15 cm/µs shown in Fig. 10. It is assessed from the numerical experiments that the efficiency of neon $Y_{sxr}$ yield in optimized

UNU/ICTP PFF (2.16%) is almost three times greater than that in optimized AECS-PF2 (0.77%) shown in Fig. 11. The difference in efficiency of these two optimized DPF machines can be explained as:

(i) The variation trend of $z_0$, '$a$' and '$b$' with $L_0$ in both machines are same.

(ii) The optimum values of '$a$' for each $L_0$ in UNU/ICTP PFF is slightly larger than that in AECS-PF2.

(iii) For each $L_0$ the optimum value of '$b$' is significantly higher in UNU/ICTP PFF than that of AECS-PF2 as shown in Fig. 12.

As the values of '$b$' is significantly large and '$a$' is slightly high in optimized UNU/ICTP PFF, the pinch column length and duration of pinch are higher than those of AECS-PF2 and $Y_{sxr}$ is proportionally related to them. As a result, the neon $Y_{sxr}$ yield, as well as its corresponding efficiency at the optimum combination of ($L_0$, $z_0$, '$a$' and '$b$') in UNU/ICTP PFF about three times higher from the optimized AECS-PF2. Therefore, in optimization of AECS-PF2, it may have computed better result with more increase in electrode radius.

Based on the obtained results of these sets of numerical experiments with neon gas, we can say that to improve the neon $Y_{sxr}$ yield, $L_0$ should be reduced to a value around 10 – 20 nH, which is an achievable range incorporating low inductance technology (NX2 machine constructed with $L_0$ = 20 nH), below which the pinch current $I_{pinch}$ and the $Y_{sxr}$ yield, as well as the corresponding efficiency would not be improved sufficiently, if at all. Moreover, the neon $Y_{sxr}$ yield may be improved twelve to thirteen times from the standard UNU/ICTP PFF by the remarkable increase in '$b$' and '$a$' with reducing $z_0$ and $L_0$, keeping c = $b/a$ constant at 3.368 in the laboratory.

**Conclusions**

The Lee model code (version: RADPFV6.1b) is applied to characterize and optimize the UNU/ICTP PFF machine operated at 14 kV and 30μF as a source of neon $Y_{sxr}$ yield. The reduction effects of $L_0$ and the electrodes geometry ($z_0$, '$a$' and '$b$') on the neon $Y_{sxr}$ yield is studied.

In our numerical experiments with this DPF machine, the neon $Y_{sxr}$ yield increases to 63.61 J at operating $P_0$ = 3.3 Torr with the corresponding efficiency of 2.16% at the optimum combination of ($L_0$ = 10 nH, $z_0$ = 2.4 cm, '$a$ = 2.0 cm' and '$b$ = 6.736 cm'). The limitation effect of $L_0$ on neon $Y_{sxr}$ yield is also observed from these numerical experiments. From the published paper, it is understood that the maximum measured neon $Y_{sxr}$ of UNU/ICTP PFF machine was (5.4 ± 1) J at optimum pressure of $P_0$ = 3.0 Torr with corresponding efficiency of 0.18%. At the present configuration of the machine, the computed maximum neon $Y_{sxr}$ yield was 3.92 J at optimum pressure of $P_0$ = 3.3 Torr with corresponding efficiency of 0.13%. On the other hand, at optimum anode configuration, neon $Y_{sxr}$ yield was computed to 9.5 J at optimum pressure of $P_0$ = 3.5 Torr with corresponding efficiency of 0.32%. Our computed value of neon $Y_{sxr}$ yield is twelve to thirteen times higher than the experimentally measured value.

This obtained result of neon $Y_{sxr}$ yield and corresponding efficiency are also compared with the computed results of AECS-PF2 operated at 15 kV and 25 µF. The values of '*a*' and '*b*' of our optimized UNU/ICTP PFF at $L_0$ = 10 nH are significantly larger than those for optimized AECS-PF2 at $L_0$ = 15 nH. This increase in electrodes radius in the optimized UNU/ICTP PFF improves the efficiency of neon $Y_{sxr}$ yield about three times from the optimized AECS-PF2.

**Acknowledgments**

The authors would like to appreciate Prof. S. Lee for his valuable support through the workshop 'NEWPF2016'. The authors are grateful to Prof. M. Akel for providing the relevant documents and information. Gratitude from the authors goes to Dr. M. A. Humayun and M. S. S. Chowdhury for their encouragements to run the research work.

Table 1: The optimum combination of $z_0$, '$a$' & '$b$' in each value of $L_0$ with their corresponding neon soft X-ray yield for UNU/ICTP PFF at fixed c = b/a = 3.368, $P_0$ = 3.3 Torr, $V_0$ = 14 kV and $C_0$ = 30 μF:

| $L_0$ | $z_0$ | $a$ | $b$ | $I_{peak}$ | $I_{pinch}$ | $I_{pinch}/I_{peak}$ | $v_a$ | $a_{min}$ | $z_{max}$ | $Y_{sxr}$ | Efficiency |
|---|---|---|---|---|---|---|---|---|---|---|---|
| nH | cm | cm | cm | kA | kA | - | cm/us | cm | cm | J | % |
| 100 | 8 | 1.25 | 4.21 | 190.92 | 139.45 | 0.730 | 4.47 | 0.08 | 1.82 | 9.88 | 0.34 |
| 75 | 7 | 1.37 | 4.614 | 217.88 | 155.93 | 0.716 | 4.65 | 0.09 | 2.00 | 14.51 | 0.49 |
| 50 | 4.85 | 1.58 | 5.321 | 260.46 | 180.96 | 0.695 | 4.70 | 0.10 | 2.34 | 24.57 | 0.84 |
| 30 | 4.3 | 1.74 | 5.86 | 318.45 | 202.68 | 0.636 | 5.27 | 0.12 | 2.63 | 37.62 | 1.28 |
| 20 | 3.8 | 1.84 | 6.197 | 368.11 | 215.43 | 0.585 | 5.78 | 0.14 | 2.84 | 48.86 | 1.66 |
| 15 | 3.39 | 1.91 | 6.44 | 405.21 | 221.13 | 0.546 | 6.12 | 0.15 | 3.03 | 57.58 | 1.96 |
| **10** | **2.4** | **2.00** | **6.736** | **453.91** | **223.94** | **0.493** | **6.42** | **0.18** | **3.23** | **63.61** | **2.16** |
| 5 | 1.3 | 2.00 | 6.736 | 508.44 | 212.95 | 0.419 | 7.03 | 0.21 | 3.25 | 55.63 | 1.89 |
| 3 | 1.27 | 2.50 | 8.42 | 623.21 | 195.21 | 0.313 | 7.12 | 0.36 | 3.90 | 19.08 | 0.65 |
| 1 | 1.26 | 2.60 | 8.757 | 729.94 | 177.86 | 0.244 | 8.47 | 0.18 | 4.05 | 8.08 | 0.27 |

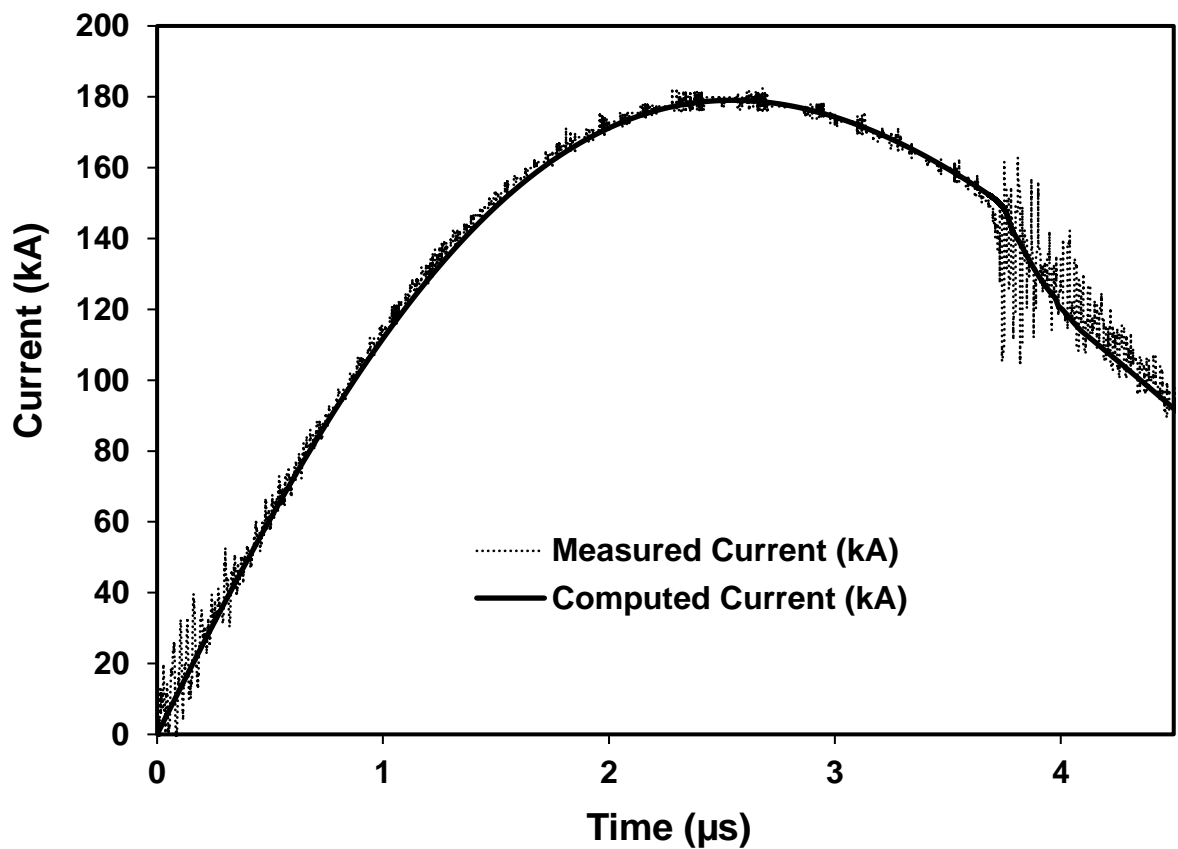

Fig. 1: Comparison of the experimental current waveform with the computed one of the UNU/ICTP PFF at 14 kV, 2.8 Torr with neon filling gas.

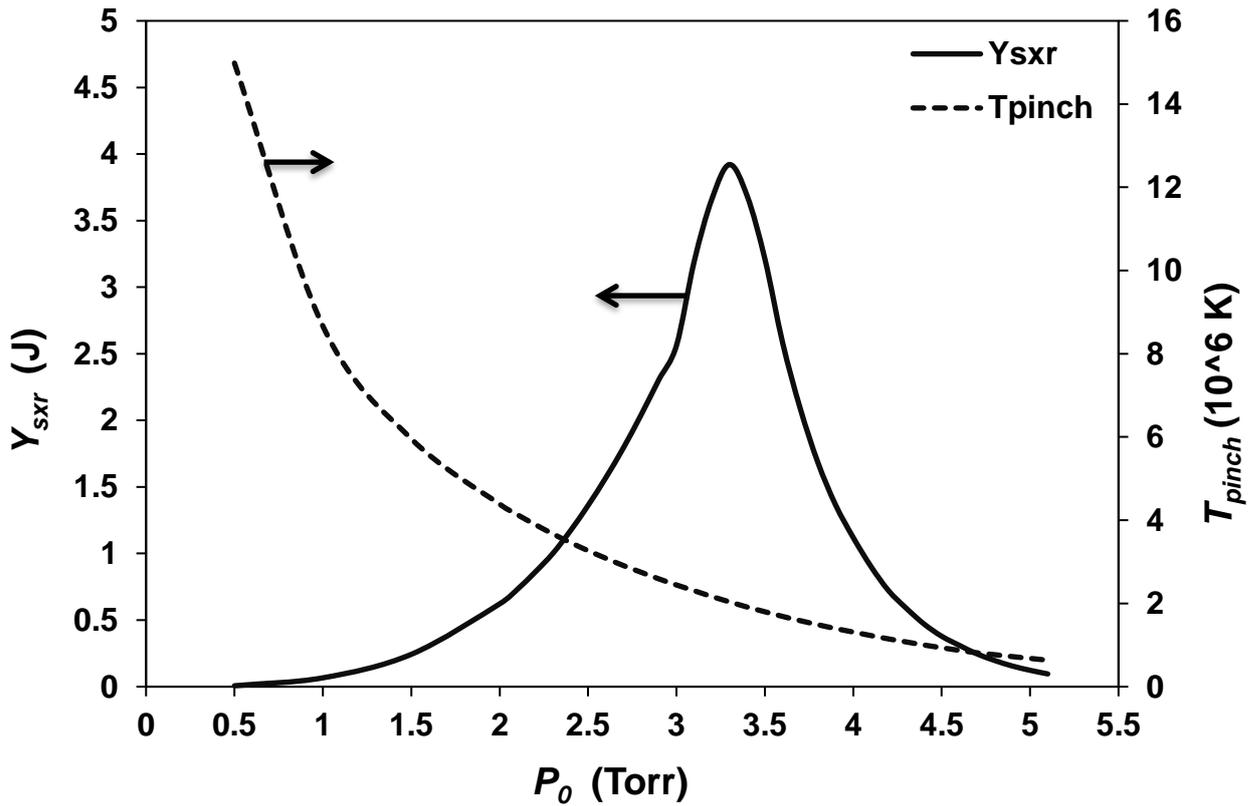

Fig. 2: Computed neon $Y_{sxr}$ yield, plasma temperature with respect to pressure of UNU/ICTP PFF at 14 kV.

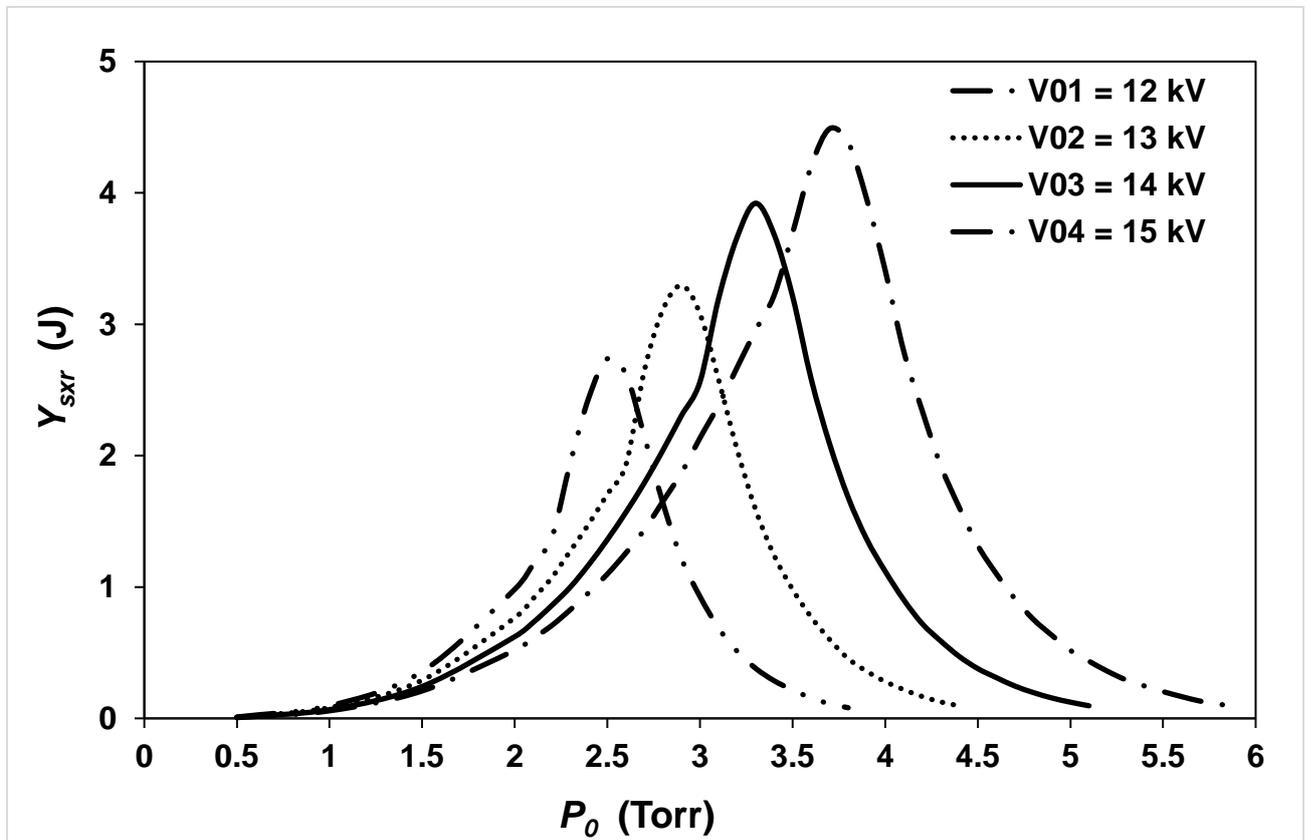

Fig. 3: Variation (computed) of neon $Y_{sxr}$ yields from UNU/ICTP PFF with pressure at different applied voltages.

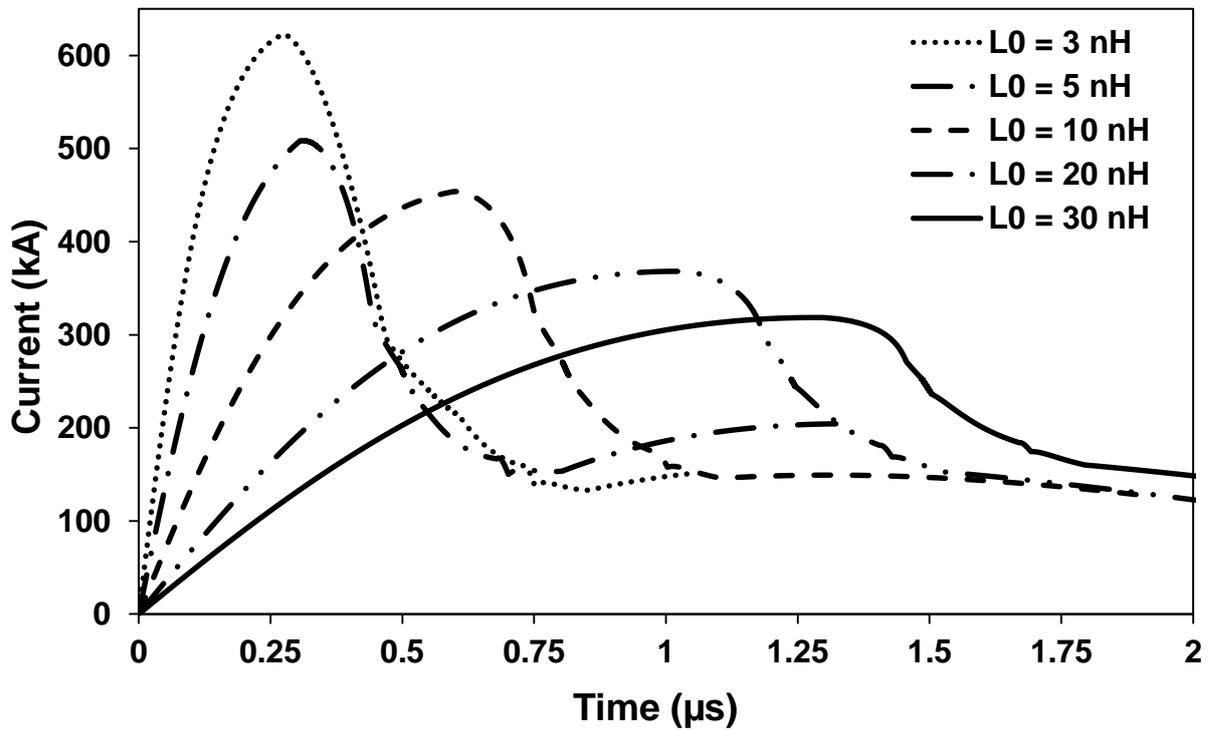

Fig. 4: Computed discharge current waveforms as a function of time from UNU/ICTP PF at 14 kV and 3.3 Torr neon with different values of $L_0$.

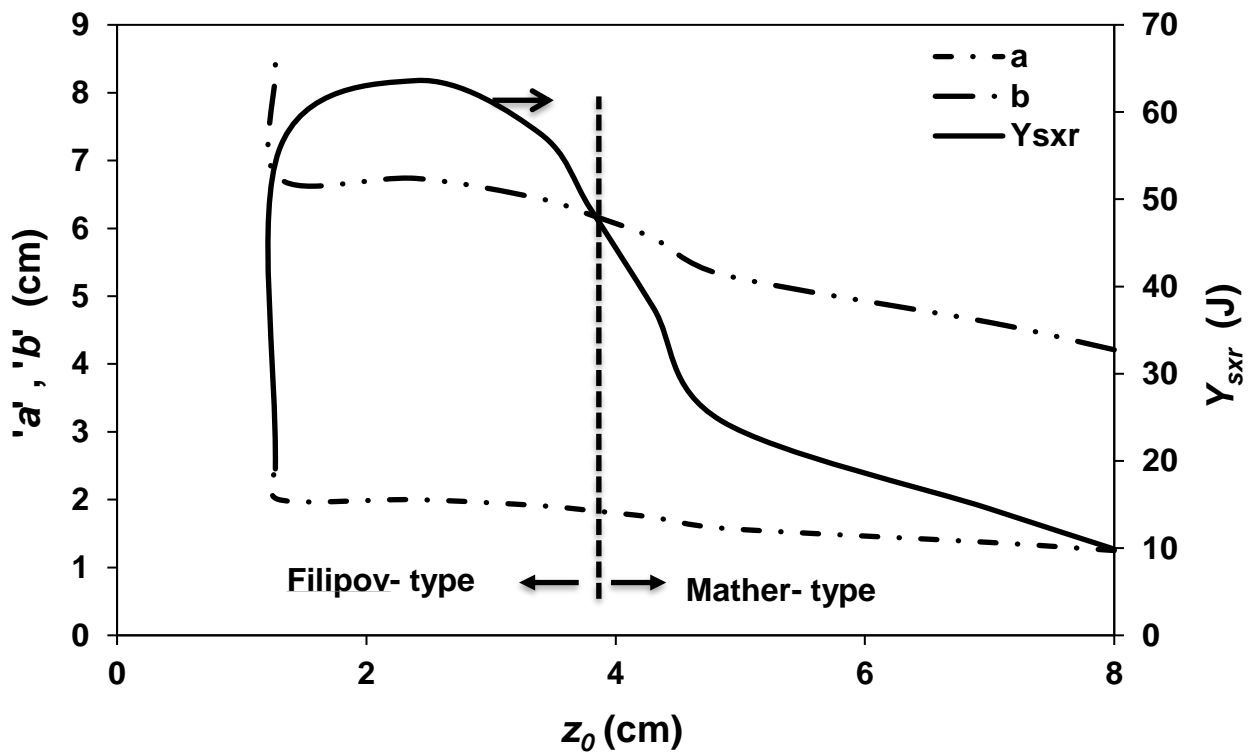

Fig. 5: Computed neon $Y_{sxr}$ yield for each optimum combination of '$a$' and '$b$' with respect to $z_0$ from UNU/ICTP PFF at 14 kV and 3.3 Torr neon.

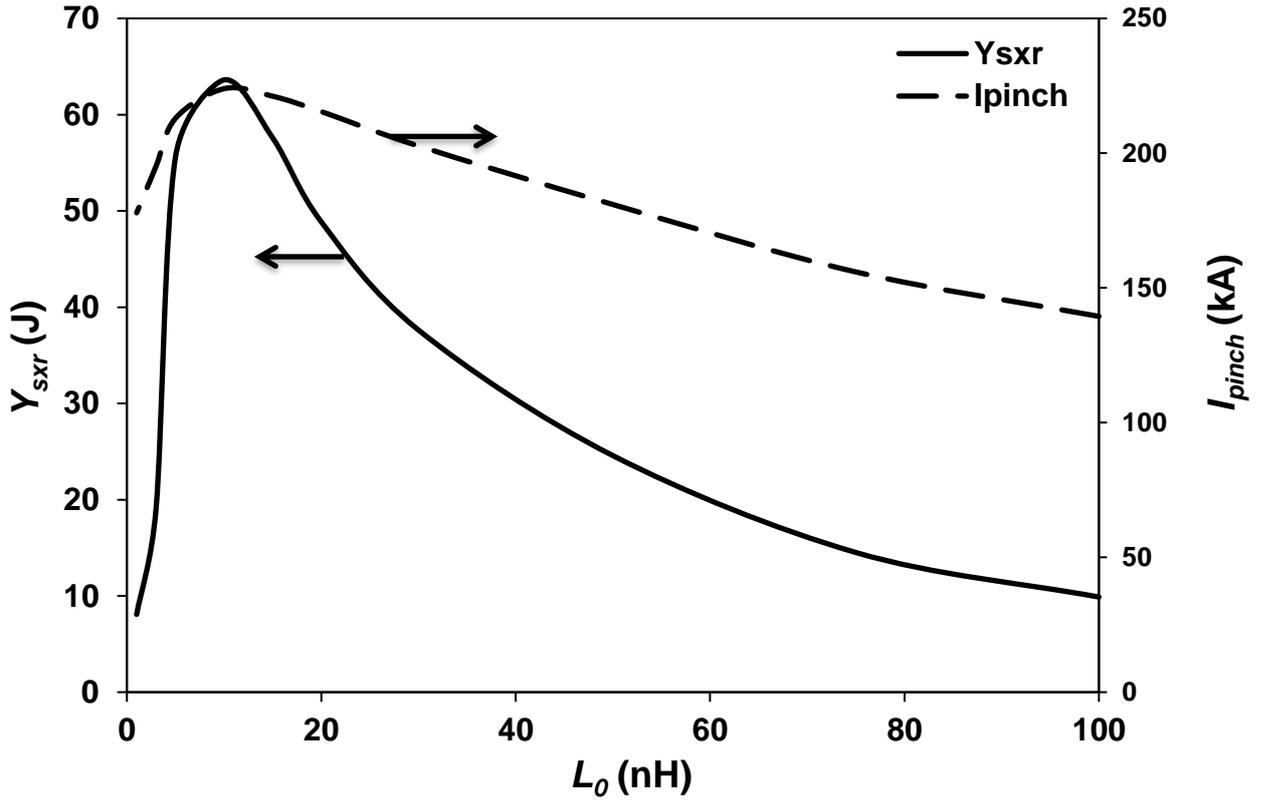

Fig. 6: Computed $Y_{sxr}$ yield and $I_{pinch}$ with respect to $L_0$ from UNU/ICTP PFF at 14 kV and 3.3 Torr neon.

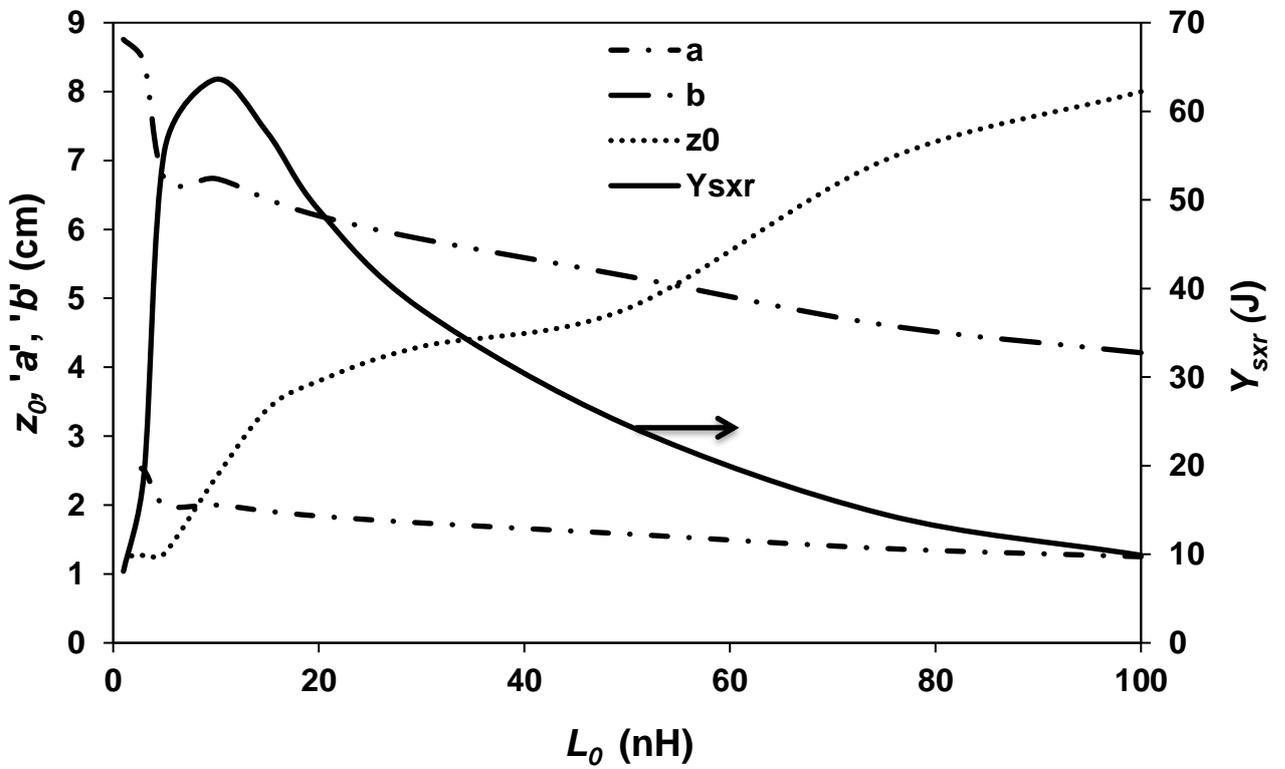

Fig. 7: Computed $Y_{sxr}$ yields and its corresponding optimum electrode's geometry with respect to $L_0$ in UNU/ICTP PFF at 14 kV and 3.3 Torr neon.

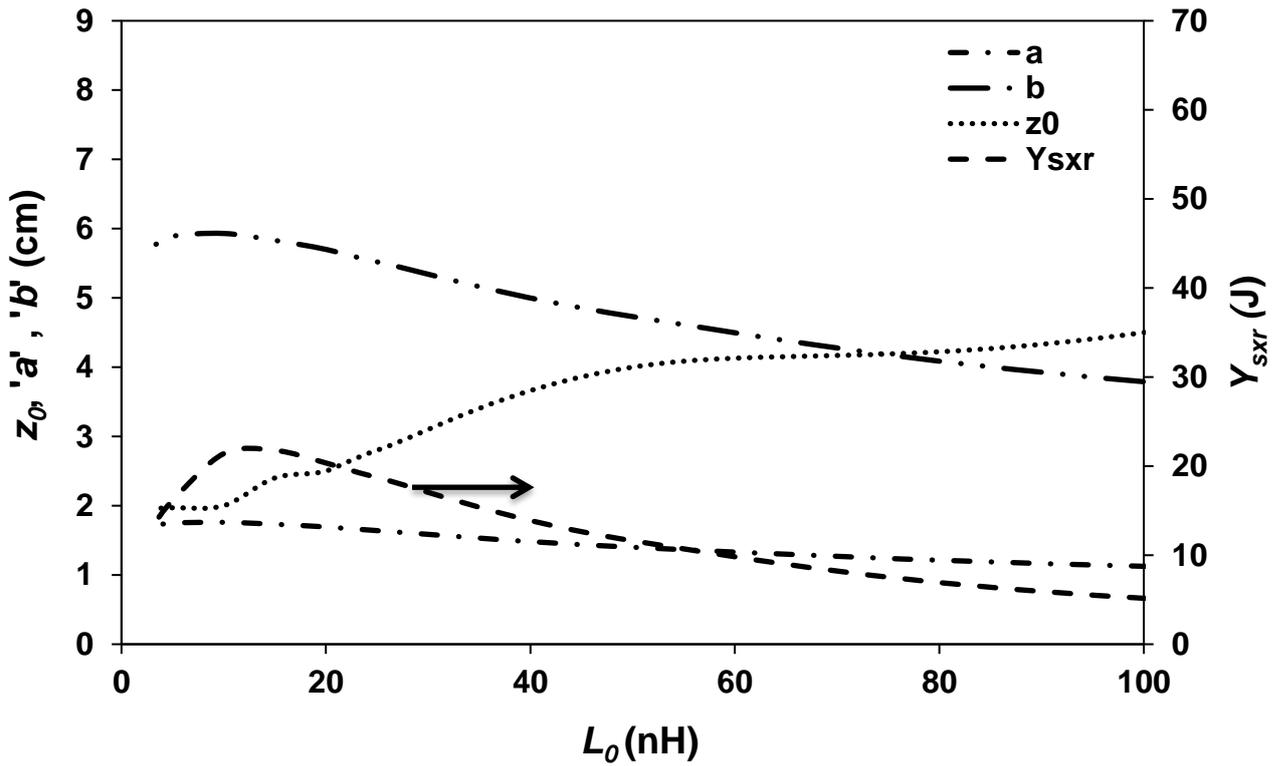

Fig. 8: Calculated $Y_{sxr}$ yields and its corresponding optimum electrode geometry with respect to $L_0$ in AECS-PF2 at 15 kV and 2.8 Torr neon.

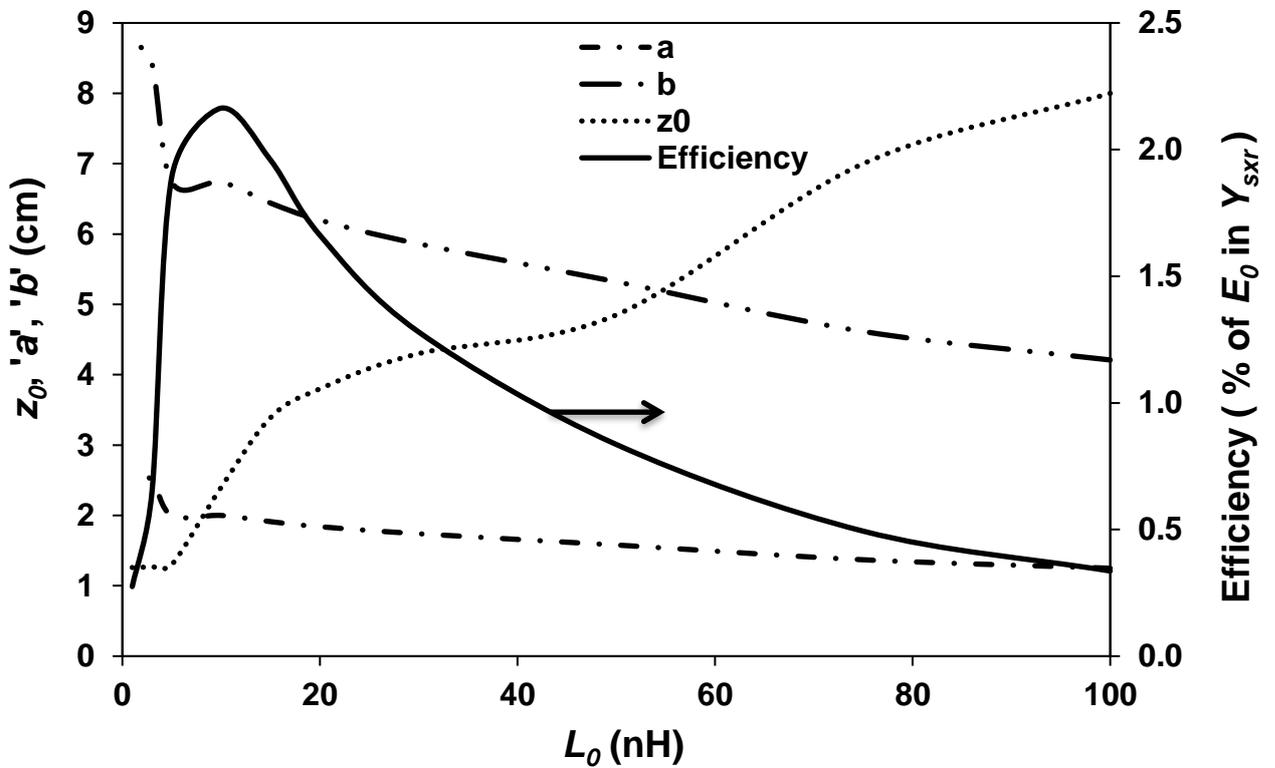

Fig. 9: Variation of efficiency (% of $E_0$ converted into $Y_{sxr}$) and its corresponding optimum values of $z_0$, '$a$' and '$b$' with $L_0$ in UNU/ICTP PFF at 14 kV and 3.3 Torr neon.

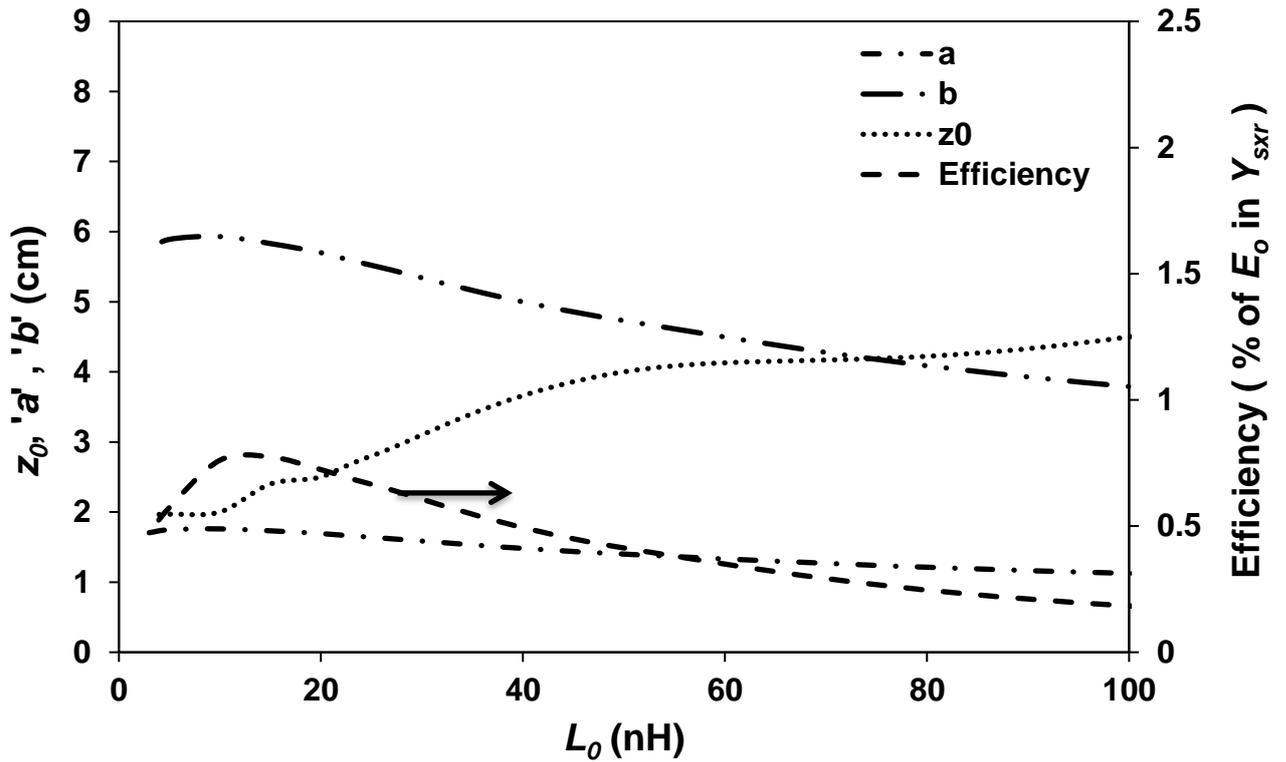

Fig. 10: Variation of efficiency and its corresponding optimum values of $z_0$, '$a$' and '$b$' with $L_0$ in AECS-PF2 at 15 kV and 2.8 Torr neon.

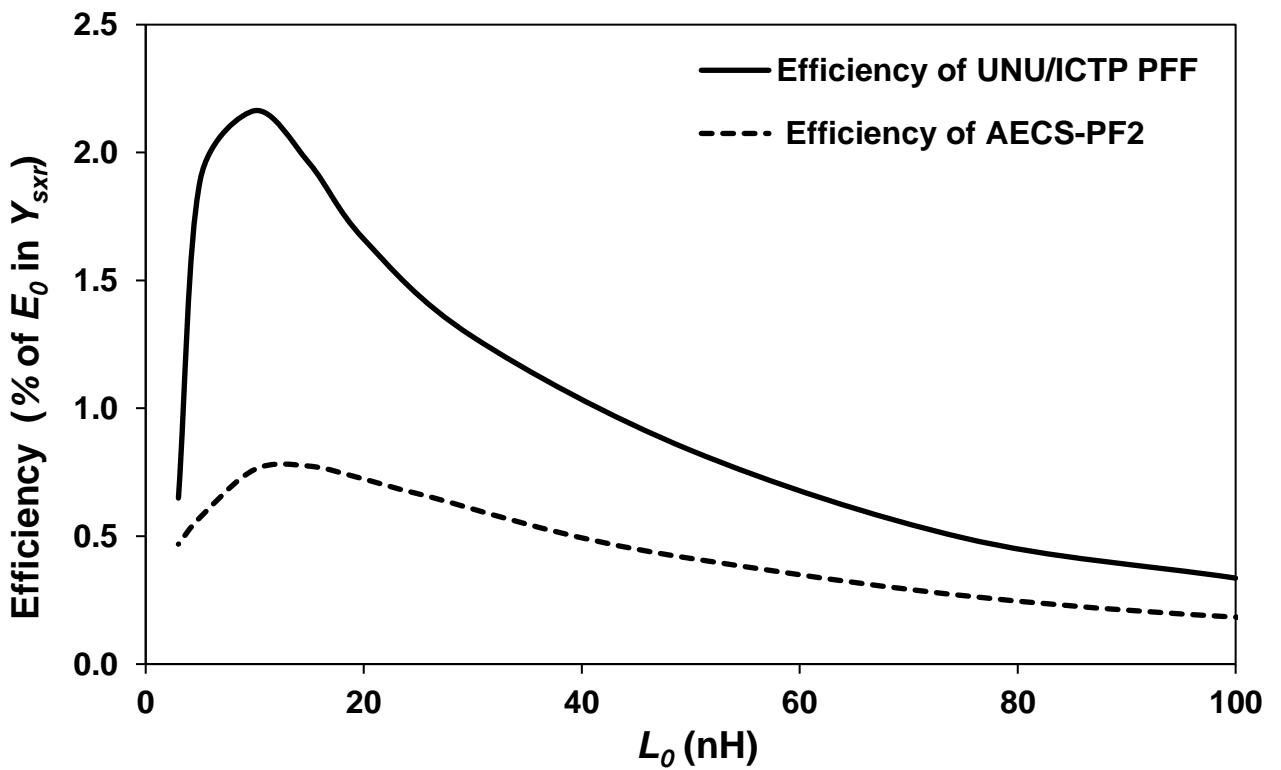

Fig. 11: Comparison of efficiency of UNU/ICTP PFF and AECS-PF2 devices at each optimum combination of electrode's geometry ($z_0$, '$a$' and '$b$') with $L_0$.

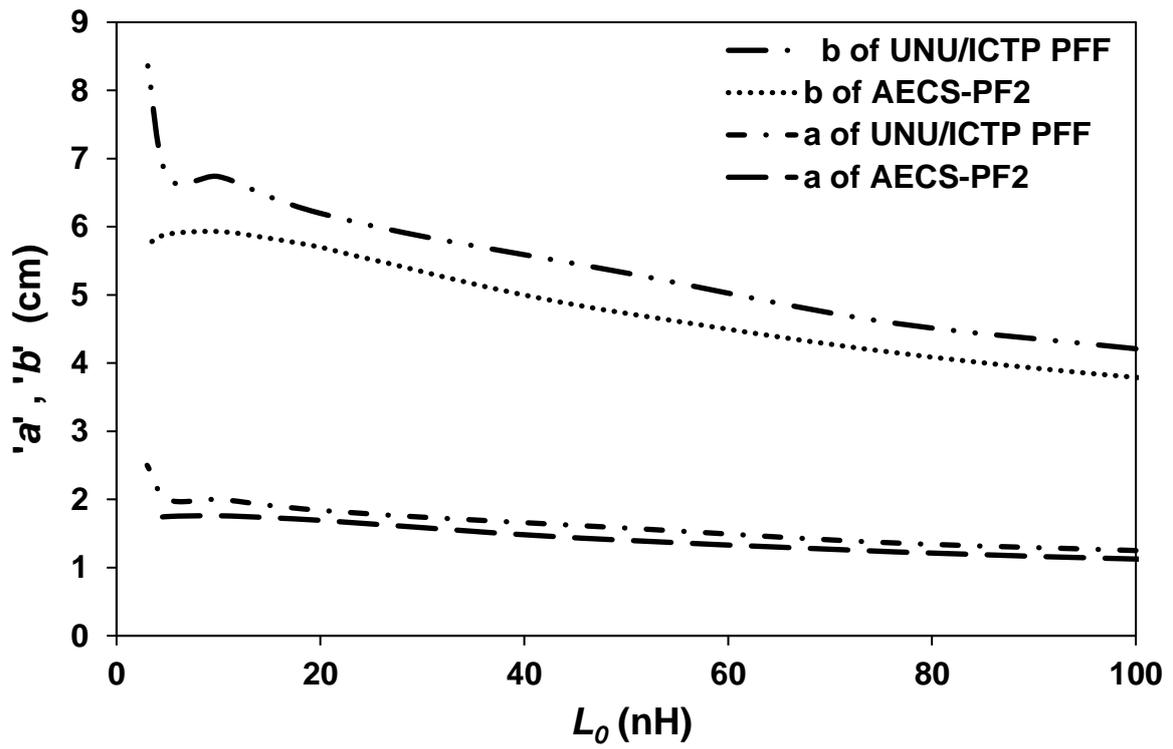

Fig. 12: Comparison of optimum electrode radius ('$a$' and '$b$') for maximum neon $Y_{sxr}$ (computed) yields with $L_0$ for UNU/ICTP PFF and AECS-PF2.